\documentclass[a4paper,aps,reprint,pra,showpacs,superscriptaddress,aps,nofootinbib]{revtex4-1}

\usepackage[utf8]{inputenc}
\usepackage[T1]{fontenc}
\usepackage[USenglish]{babel}
\usepackage[sc,osf]{mathpazo}
\usepackage[scaled=0.86]{berasans}
\usepackage[scaled=1.03]{inconsolata}
\usepackage[colorlinks]{hyperref}
\usepackage{graphicx}
\usepackage{amsmath,amssymb,amsthm}
\usepackage{xspace}
\usepackage{mathtools}
\usepackage{amsmath}

\let\originalleft\left
\let\originalright\right
\renewcommand{\left}{\mathopen{}\mathclose\bgroup\originalleft}
\renewcommand{\right}{\aftergroup\egroup\originalright}

\newcommand{\bra}[1]{\left\langle #1 \right|}
\newcommand{\ket}[1]{\left| #1 \right\rangle}

\newcommand{\kett}[1]{| #1 \rangle}

\newcommand{\norm}[1]{\left\|#1\right\|}

\newcommand{\de}[1]{\left(#1\right)}

\newcommand{\id}{\mathbb{1}}

\newcommand{\eg}{\emph{e.g.}\@\xspace}
\newcommand{\ie}{\emph{i.e.}\@\xspace}

\mathtoolsset{centercolon}

\input{Qcircuit}

\hypersetup{pdftitle={{Computational advantage from quantum-controlled ordering of gates}}}

\begin{document}

\title{Computational advantage from quantum-controlled ordering of gates}

\author{Mateus Araújo}
\author{Fabio Costa}
\author{Časlav Brukner}
\affiliation{Faculty of Physics, University of Vienna, Boltzmanngasse 5, 1090 Vienna, Austria}
\affiliation{Institute for Quantum Optics and Quantum Information (IQOQI), Austrian Academy of Sciences, Boltzmanngasse 3, 1090 Vienna, Austria}


\date{\today}


\begin{abstract}
It is usually assumed that a quantum computation is performed by applying gates in a specific order. One can relax this assumption by allowing a control quantum system to switch the order in which the gates are applied. This provides a more general kind of quantum computing, that allows transformations on blackbox quantum gates that are impossible in a circuit with fixed order. Here we show that this model of quantum computing is physically realizable, by proposing an interferometric setup that can implement such a quantum control of the order between the gates. We show that this new resource provides a reduction in computational complexity: we propose a problem that can be solved using $O(n)$ blackbox queries, whereas the best known quantum algorithm with fixed order between the gates requires $O(n^2)$ queries. Furthermore, we conjecture that solving this problem in a classical computer takes exponential time, which may be of independent interest.
\end{abstract}



\maketitle

\textit{Introduction.}---A useful tool to calculate the complexity of a quantum algorithm is the blackbox model of quantum computation. In this model, the input to the computation is encoded in a unitary gate -- treated as a blackbox -- and the complexity of the algorithm is the number of times this gate has to be queried to solve the problem. 

Typically, black-box computation is studied within the quantum circuit formalism~\cite{deutsch89}. A quantum circuit consists of a collection of wires, representing quantum systems, that connect boxes, representing unitary transformations. In this framework, wires are assumed to connect the various gates in a fixed structure, thus the order in which the gates are applied is determined in advance and independently of the input states. It was first proposed in \cite{chiribella09a} that such a constraint can be relaxed: one can consider situations where the wires, and thus the order between gates, can be controlled by some extra variable. This is natural if one thinks of the circuit's wires as quantum systems that can be in superposition.

Such ``superpositions of orders'' allow performing information-theoretical tasks that are impossible in the quantum circuit model: it was shown in \cite{chiribella12} that it is possible to decide whether a pair of blackbox unitaries commute or anticommute with a single use of each unitary, whereas in a circuit with a fixed order at least one of the unitaries must be used twice. (The same task was considered in a quantum optics context in \cite{andersson05}, where a less efficient protocol was found.)

It was not known, however, whether this advantage can be translated into more efficient algorithms for quantum computing, \ie, if a quantum computer that can control the order between gates can solve a computational problem with asymptotically less resources than a quantum computer with fixed circuit structure.

Here we present such a problem: given a set of $n$ unitary matrices and the promise that they satisfy one out of $n!$ specific properties, find which property is satisfied. The essential resource to solve this problem is the quantum control over the order of $n$ blackboxes, first introduced in Ref.~\cite{colnaghi11}. We show that, by using this resource, the problem can be solved with $O(n)$ queries to the blackboxes, while the best known algorithm with fixed order requires $O(n^2)$ queries. Furthermore, while both quantum methods of solving the problem run in polynomial time, the best known classical algorithm to solve it runs in exponential time, which may be of independent interest.

We further discuss a possible interferometric implementation of the protocol. For the superposition of the order of just two gates, a realization with current quantum optics techniques is possible. For a higher number of gates practical implementations become more challenging.

\textit{Algorithm.}---The quantum control of the order between $n$ unitary gates can be formalized by introducing the $n$-\textsc{switch} gate. As in Ref.~\cite{colnaghi11}, we consider a $d$-dimensional target system, initialized in some state $\ket{\psi}$, and an $n!$-dimensional control system. Let $\{U_i\}_0^{n-1}$ be a set of unitaries and
\begin{equation}
\label{product}
 \Pi_x = U_{\sigma_x(n-1)}\ldots U_{\sigma_x(1)}U_{\sigma_x(0)}
\end{equation}
	for some permutation $\sigma_x$, where $x=0,\dots, {n!-1}$ is a chosen labelling of permutations\footnote{More preciselly, $\sigma_x(j)$ is the image of the $j$th element under the permutation $x$}. Then the $n$-\textsc{switch} $S_n$ is a controlled quantum gate: its effect is to apply the product of unitaries $\Pi_x$ to the state $\ket{\psi}$ conditioned on the value of the control register $\ket{x}$. In symbols,
\begin{equation}
	\label{n-switch}
	S_n\ket{x}\ket{\psi}=\ket{x}\Pi_x\ket{\psi}. 
\end{equation} 

Using this gate we can introduce an algorithm that exploits the quantum control of orders to achieve a reduction in query complexity for the solution of a specific problem. The algorithm is based on the standard Hadamard test. The idea is to initiate the control system in a state corresponding to a uniform superposition of all permutations, apply $S_n$, and then measure the control system in the Fourier basis. With a suitable choice of the unitaries, we can make the result of this measurement deterministic and, since there are $n!$ different results, this means that we can differentiate between $n!$ different properties of $n$ unitaries.

	To be more precise, let $\omega = e^{i\frac{2\pi}{n!}}$. We say that the set of unitaries $\{U_i\}_0^{n-1}$ has property $\textbf{P}_y$ if it is true that
	\begin{equation}\label{eq:property}
	\forall x\ \Pi_x=\omega^{xy}\Pi_0,
	\end{equation}
	for the given $y$. For example, property $\textbf{P}_0$ is the property that $\Pi_x = \Pi_0$ for all $x$, \ie, that all the matrices commute with each other. 
	
	Note that it is not possible to satisfy property $\textbf{P}_1$	if the dimension of the unitaries $d$ is less than $n!$. To see that, consider $x=y=1$, and take the determinant on both sides of equation \eqref{eq:property}:
\begin{equation}
\label{determinant}
\det \Pi_1 = \omega^d \det \Pi_0. 
\end{equation}
 	Since $\det \Pi_x = \det \Pi_0$, it follows that $\omega^d =1$, and therefore $d$ must be at least $n!$.
	
	The computational problem is defined as follows: given a set $\{U_i\}_0^{n-1}$ of unitary matrices of dimension $d\ge n!$, decide which of the properties $\textbf{P}_y$ is satisfied by this set, given the promise that one of these $n!$ properties is satisfied. 
	
	The protocol for solving this problem is the following: we initialize the target system in \textit{any} state $\ket{\psi}$, and the control system in the state $\ket{C}$ which corresponds to an equal superposition of all permutations:
	\begin{equation}
		\label{initial}
		\ket{C}\ket{\psi} = \frac{1}{\sqrt{n!}}\sum_{x=0}^{n!-1}\ket{x}\ket{\psi}.
	\end{equation}
	Then, we apply the $n$-\textsc{switch}:
	\begin{equation}
		\label{step1}
		S_n\ket{C}\ket{\psi} = \frac{1}{\sqrt{n!}}\sum_{x=0}^{n!-1}\ket{x}\Pi_x\ket{\psi}.
	\end{equation}
	Now we apply the Fourier transform over $\textbf{Z}_{n!}$ to our control qudit
	\begin{equation}
		\label{step2}
		\mathcal{F}_{n!}S_n\ket{C}\ket{\psi} = \frac{1}{n!}\sum_{x,s=0}^{n!-1}\ket{s}\omega^{-xs}\Pi_x\ket{\psi}.
	\end{equation}
	and measure the control qudit in the computational basis, with outcome probabilities
	\begin{equation}
		p_s = \frac{1}{n!^2}\norm{\sum_{x=0}^{n!-1}\omega^{-xs}\Pi_x\ket{\psi}}^2.
	\end{equation}
	Using the promise $\Pi_x=\omega^{xy}\Pi_0$ we get that
	\begin{equation}
		p_s = \frac{1}{n!^2}\norm{\sum_{x=0}^{n!-1}\omega^{x(y-s)}\Pi_0\ket{\psi}}^2 = \delta_{sy},
	\end{equation}
	that is, if property $\textbf{P}_y$ is true, the result of the measurement is going to be $y$ with probability one, so we can find out which property the unitaries have in a single run of the protocol.
	
	We should notice that the problem is not trivial, \ie there exist, for every $n$, infinitely many sets of unitary matrices that satisfy each of the $n!$ properties $\textbf{P}_y$ (see Appendix \ref{sec:fabio}) The problem, and the corresponding protocol, can be also modified to tolerate possible experimental error. This modification is shown in Appendix \ref{sec:tolerating}.

	\textit{Query Complexity.}--We are interested in determining the number of times that the unitaries $U_i$ must be used to run the algorithm. Clearly this depends only on the implementation of the $n$-\textsc{switch} gate, since the unitaries are not used anywhere else. As proposed in \cite{chiribella09a}, the \textsc{switch} can in principle be implemented by adding quantum control to the connections between the unitaries. In such an implementation it is sufficient to use a single copy of each unitary, while the control system determines the order in which the target system passes through the unitaries.

	Since the implementation with quantum control of the connections between gates is explicitly outside the quantum circuit formalism, we cannot simply calculate the number of uses of the unitaries by counting the number of times they appear in a circuit. Nevertheless, we can formulate the notion of ``gate uses'' in a precise, operational, way. Imagine we append, to each gate, an additional ``flag'' quantum system that counts the number of times that gate is used. This can be done in a reversible way: the $j$-th flag is initialized in the state $\ket{0}_{j}$ and, whenever the unitary $U_j$ is used, it is updated through the unitary transformation $\ket{f}_{j}\rightarrow\ket{f+1}_{j}$. It is easy to see that, after applying the $n$-\textsc{switch}, the state of the flags factorizes, with each flag in the state $\ket{1}_j$. According to this definition, the total number of queries necessary to run the algorithm is $n$.
	
	In comparison, the optimal simulation of the $n$-\textsc{switch} gate with a fixed circuit has query complexity $\Omega(n^2)$ \footnote{We say that $f(n)$ is $\Omega(g(n))$ if there exists a constant $M$ such that $f(n) \ge M g(n)$ for sufficiently large $n$.}. To see this, first note that one can assume without loss of generality that all blackbox unitaries are applied each in a different time step, since if two blackboxes are applied in parallel, we can always introduce a time delay between them, without changing the action of the circuit. More technically, a circuit defines a partial order for its gates, which can always be completed into a total order. Then, let $\{A_i\}_0^{m-1}$ be the $m$ blackbox unitaries appearing in the circuit, with $A_j \in \{U_i\}_0^{n-1}$, queried in the order $A_0 \preceq\dots\preceq A_{m-1}$. From the Appendix B of Ref.~\cite{chiribella09a}, it follows that it is only possible to apply the $\Pi_x$ to $\ket{\psi}$ if the unitaries $U_{\sigma_x(0)},\dots,U_{\sigma_x(n-1)}$ are present in the circuit in the order defined by $\sigma_x$. The lower bound on the query complexity is then the minimal $m$ for which all $n!$ permutations of $\{U_0,\ldots,U_{n-1}\}$ are present as subsequences of the sequence $\{A_0,\ldots,A_{m-1}\}$.
	
	It turns out that this is an open problem in combinatorics \cite{kleitman76,radomirovic12}. However, it is known that the optimal $m$ respects the bounds
	\[ n^2 - C_\epsilon n^{7/4+\epsilon} \le m \le \left\lceil n^2 - \frac73n + \frac{19}3 \right\rceil \]	
	for any $\epsilon > 0$, where $C_\epsilon$ is a constant that depends on $\epsilon$. This concludes the proof.
	
	It is also possible to construct a quantum circuit that simulates the $n$-\textsc{switch} gate from such a sequence. We shall, however, refrain from doing so. Instead, for completeness, we present a simple circuit that simulates the $n$-\textsc{switch} gate using $m = n^2$ queries in the Appendix \ref{sec:switchbqp}.
	
	Of course, it might not be necessary to use the $n$-\textsc{switch} gate in order to determine which property $\textbf{P}_y$ the unitaries satisfy. For example, it is possible to solve the problem by directly measuring the phase obtained when applying the permutation $\sigma_1$. Since $\Pi_1 = \omega^y \Pi_0$, this is sufficient to determine $y$. However, this protocol can work only if the relative phase is measured with an error smaller than $\frac{2\pi}{n!}$, and for blackbox unitaries this can only be done with an exponential amount of queries. 
	
	This is the case, for example, for Kitaev's phase estimation algorithm \cite{kitaev95}. This algorithm is not usually applied to blackbox unitaries, but this can be done using the techniques in \cite{zhou11,zhou13,araujo13a}. In this case, to calculate the phase with the required $O(n\log n)$ bits of precision, one would need to implement the matrices controlled-$U_i^{2^{k}}$, with $k=1,\dots,n\log n$, which would require an exponential amount of queries to the blackboxes $U_i$. Even if one assumes that it is possible to apply controlled-$U_i^{2^{k}}$ efficiently -- a necessary assumption to make Kitaev's algorithm efficient -- one would need $O(n\log n)$ queries to each $U_i^{2^{k}}$ oracle. Since there are $n$ unitaries $U_i$, the query complexity would be $O(n^2\log n)$,  which is still less efficient than simulating the $n$-\textsc{switch} with a fixed circuit.
	
	\textit{Running time.}---Instead of query complexity we may want to consider the running time of the algorithm. If we assume that this is dominated by applying the unitaries $U_i$, then there is no difference between the implementation with superposition of orders or the fixed quantum circuit: both run in time $O(n)$ (see Appendix \ref{sec:switchbqp}).
	
	It is interesting, nevertheless, to compare the time required to solve the problem between quantum and classical computers. If we assume that the unitaries $U_i$ are decomposed in a polynomial amount of elementary gates, they can be given as an input of polynomial size to a classical algorithm, and it makes sense to compare the classical and quantum running times.
	
	As argued above, the problem of determining $y$ reduces to the problem of calculating the relative phase between $\Pi_1$ and $\Pi_0$, which may differ by the permutation of a single pair of unitaries. However, as discussed before, the dimension of the unitaries must be at least $n!$ for this problem to be nontrivial, and it seems unlikely that one could extract the phase from these exponentially large unitary matrices on a classical computer in polynomial time. On the other hand, the running time of the quantum algorithm is clearly polynomial for unitaries decomposed in a polynomial amount of elementary gates. Therefore we conjecture that for the problem presented there is an exponential separation between classical and quantum complexity, which may be of independent interest.

	Note that our algorithm is based on the quantum Fourier transform, as are several algorithms that show an exponential separation between classical and quantum complexity, but there does not appear to be a more direct connection with specific classes of quantum algorithms, such as those that solve the hidden subgroup problem (see Appendix \ref{sec:other_algorithms}).
	
	\textit{Physical implementation.}---In Ref.~\cite{chiribella09a} it was proposed to apply the superposition principle to the physical components of a quantum computer that determine the order between gates. Since this requires a quantum control over macroscopic systems, it seems outside of the reach of current technology and could be practically unfeasible. Here we propose an implementation of the $n$-\textsc{switch} that, although experimentally challenging, might be feasible.

\begin{figure}[tb]
	\includegraphics[width=\columnwidth]{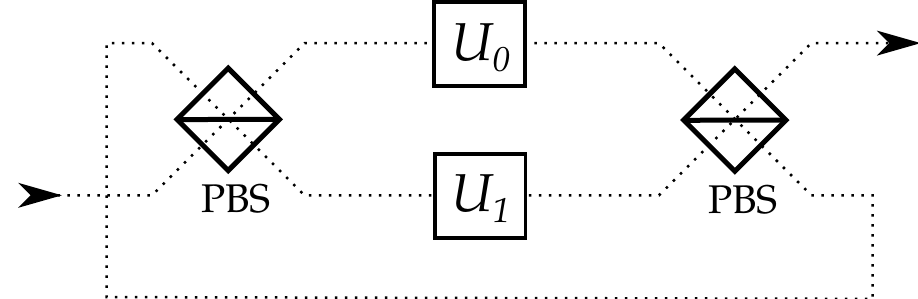}
	\caption{Linear optical implementation of the 2-\textsc{switch}. The unitaries $U_0$ and $U_1$ act on internal degrees of freedom of a single photon, such as space bins, time bins, or angular momentum modes. The polarization state of the photon determines the order in which the unitaries are applied. A photon with polarization $\ket{H}$ is transmitted by the polarizing beam splitters (PBSs), so that $U_0$ is applied before $U_1$. For a photon with polarization $\ket{V}$, reflected by the PBSs, $U_1$ is applied first and $U_0$ second.}
	\label{fig:interferometer}
	\end{figure}

We first consider an optical implementation of a 2-\textsc{switch} for $2 \times 2$ unitaries, illustrated in Fig.~\ref{fig:interferometer} (this implementation was independently developed in \cite{radu}). The control system is the polarization of a photon and the target system some internal degree of freedom of the same photon, such as space bins, time bins, or angular momentum modes. If the photon is prepared in a horizontally polarized state $\ket{H}$, it is transmitted by both polarizing beam splitters (PBSs), resulting in the application of the unitary $U_0$ first and of $U_1$ second. A photon in a vertically polarized state $\ket{V}$ is reflected by both PBSs, thus the two unitaries are applied in the reversed order. For an arbitrary polarization state $\alpha \ket{H} + \beta \ket{V}$, the photon exits the interferometer in the state $\alpha \ket{H}U_1U_0\ket{\psi} + \beta \ket{V} U_0U_1\ket{\psi}$, which corresponds to the output of the 2-\textsc{switch}.

\begin{figure}[tb]
	\includegraphics[width=0.9\columnwidth]{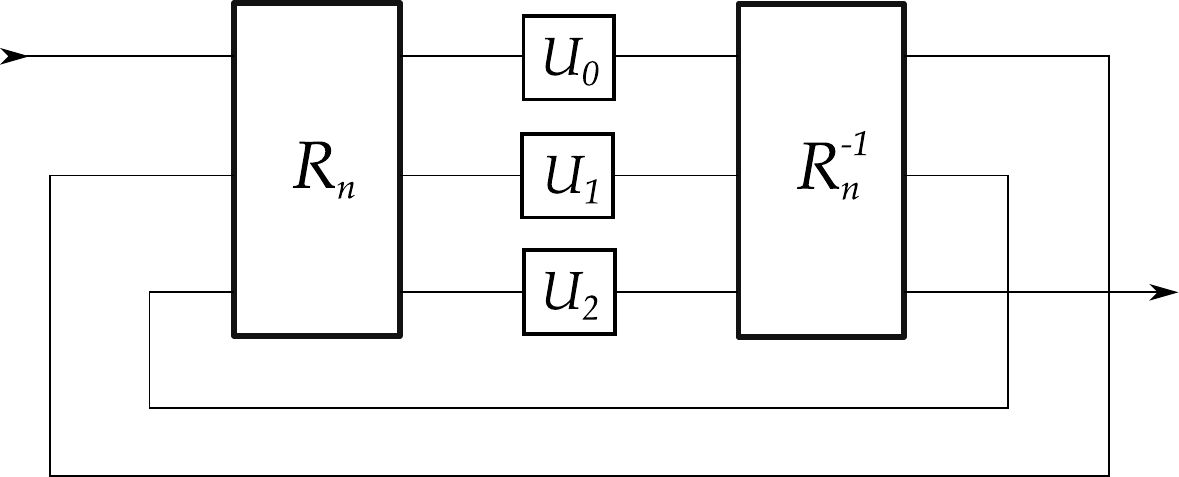}
	\caption{Implementation of the n-\textsc{switch} (in the figure, $n=3$). Both control and target, in state $\ket{x}$ and $\ket{\psi}$ respectively, are encoded in a single (possibly multi-particle) system. When the system enters the $n$-\textsc{router}s ($R_n$) in mode $j$, it is redirected to mode $\sigma_{x}(j)$ and the unitary $U_{\sigma_{x}(j)}$ is applied to $\ket{\psi}$. $R_n^{-1}$ performs the inverse permutation and sends the system to mode $j+1$ of the first router. In this way, a system entering mode $0$ of the first router, eventually exits mode $n-1$ of the second router with target in the state $\Pi_x\ket{\psi}$.}
	\label{fig:router}
	\end{figure}

The extension of this scheme to the general case of an $n$-\textsc{switch} can be obtained with a generalization of the PBS to an element, which we call $n$-\textsc{router}, with $n$ input modes and $n$ output modes (see Fig.~\ref{fig:router}). If the control system is in a state $\ket{x}$, the $n$-\textsc{router} sends the input mode $j$ to the output mode $\sigma_x(j)$. The unitary $U_{\sigma_x(j)}$ is applied to a system in the mode $\sigma_x(j)$, which then enters a second router that performs the inverse permutation. The output mode $j$ of the second router is then directed to the input mode $j+1$ of the first one. It is straightforward to check that a system entering mode $0$ of the first router in the state $\ket{x}\ket{\psi}$ exits mode $n-1$ of the second router in the state $\ket{x}\Pi_x\ket{\psi}$. (In Appendix \ref{sec:router} we show how to construct an $n$-\textsc{router} with $O(n^2)$ binary routers.)

This higher-dimensional routing can be achieved, for example, with orbital angular momentum of light \cite{mirhosseini13,berkhout10}. However, the main limitation of an optical implementation of the $n$-\textsc{router} is that it is not scalable in an obvious way, since it requires encoding an exponential number of degrees of freedom in a single photon ($n!$ for the control system and, as argued before, at least $n!$ for the target system). A scalable implementation could be obtained by encoding the degrees of freedom in $O(n\log n)$ particles, each carrying a constant number of degrees of freedom (\eg, one qubit each). The main challenge is then to implement a router that, conditioned on the multiparticle state, coherently directs all the particles in a specific mode. This is in principle possible if the particles are bound together, \eg as atoms in a molecule. Recent progress in matter-wave interferometry suggests that such a quantum control of composite systems could be achievable in the future \cite{arndt99, eibenberger13}.

Other realizations of superposition of orders, based on different models of computation, could also be possible. For example, an implementation of the 2-\textsc{switch} within adiabatic quantum computing was proposed recently~\cite{nakago13}.

\textit{Conclusion.}---We have shown that extending the quantum circuit model by allowing quantum control of the order between gates provides a reduction in the number of queries needed to solve a computational problem. Furthermore, we have proposed a physically realizable experimental scheme to implement such a control.

While the reduction is only polynomial, and thus does not create a new complexity class, the result shows that extending the quantum circuit model is possible and can provide a computational advantage. Besides, the computational problem introduced has no known efficient solution by a classical algorithm, which may be of independent interest.

Other extensions of the quantum circuit model of blackbox computation have been proposed~\cite{araujo13a}: it was shown recently~\cite{soeda13,araujo13a,thompson13} that a quantum circuit cannot apply blackbox gates conditioned on the state of a control qubit. However, such a control is physically realizable~\cite{zhou11,zhou13} and therefore should be allowed by the formalism. It is also intriguing to ask what computational advantages might be achieved once the restrictions imposed by the fixed causal structure of quantum mechanics are relaxed~\cite{oreshkov12}.

\begin{acknowledgments}
This work was supported by the Austrian Science Fund (FWF) (Project W1210 Complex Quantum Systems (CoQuS), Special Research Program Foundations and Applications of Quantum Science (FoQuS), and Individual project 24621), the European Commission Project RAQUEL, FQXi, and by the John Templeton Foundation.
\end{acknowledgments}

\bibliographystyle{linksen}
\bibliography{biblio}

\appendix

\section{Existence proof of the sets of unitaries}\label{sec:fabio}

	We need to show that for every $y$ there exists a set of unitaries that satisfies property $\textbf{P}_y$, otherwise the problem becomes trivial. More specifically, we will show that for every integer $n\geq 1$ and every $y\in \{0,\dots,n!-1\}$ it is possible to find a set of $n$ unitary matrices $\{U_i\}_0^{n-1}$ such that
\begin{equation}\label{eq:property_appendix}
\Pi_x =\omega^{xy} \Pi_0
\end{equation}
is true for all $x$ for some choice of $y$, where 
\begin{equation}
 \Pi_x = U_{\sigma_x(n-1)}\ldots U_{\sigma_x(1)}U_{\sigma_x(0)},
\end{equation}
	and $\omega = e^{i\frac{2\pi}{n!}}$.	

	We first present the construction for $y=1$. In this case, to each permutation of the matrices corresponds a different phase $\omega^x$. We start by proving that the pairwise relations
\begin{equation}
\label{eq:pairwise}
U_j U_k = \omega^{k!} U_k U_j \quad \text{for}\quad j < k
\end{equation} 
generate all the $n!$ required phases. In order to prove this, it is convenient to introduce an explicit labeling of the permutations. A generic permutation of the $n$ matrices $\{U_{n-1},\dots, U_0\}$ is labeled by a sequence of integers $\left(a_{n-1},\dots,a_1\right)$, with $0\leq a_k \leq k$, and is obtained by shifting the matrix $U_k$ to the right $a_k$ times, starting from $k=1$. As an example, let us construct the four-element permutation labeled by $\left(3,1,1\right)$. First, we swap $U_0$ and $U_1$ (\ie, we shift $U_1$ one position to the right), obtaining $U_3U_2U_0U_1$, then we shift $U_2$ one position to the right and obtain $U_3U_0U_2U_1$. Finally, we shift $U_3$ three times to the right and get the desired permutation, $U_0U_2U_1U_3$. In this procedure, each matrix $U_k$ is swapped $a_k$ times with matrices $U_j$ having $j<k$. Thus, if the relations \eqref{eq:pairwise} hold, the permutation labeled by $\left(a_{n-1},\dots,a_1\right)$ produces a phase $\omega^x$, where
\begin{equation}
\label{factoradic}
x=\sum_{k=1}^{n-1}a_k k!
\end{equation}
is the expansion of the integer $x$ in the \emph{factorial basis} (or \textit{factoradic} expansion). For the given example, $x=3\times 3! + 1\times 2!+1\times 1! = 21$.

We construct now a set of $n$ unitary matrices that satisfy the pairwise relations \eqref{eq:pairwise}. They are constructed from the generalized $X$ and $Z$ matrices
\begin{subequations}\label{eq:weylheisenberg}
\begin{gather}
X= \sum_{j=0}^{n!-1} \ket{j\oplus 1}\bra{j}, \\
Z= \sum_{j=0}^{n!-1} \omega^j \ket{j}\bra{j},
\end{gather}
\end{subequations}
where $\oplus$ denotes the sum modulo $n!$. Note that they satisfy the commutation relation
\begin{equation}\label{eq:phase}
 ZX = \omega XZ.
\end{equation}
Then we define
\begin{equation}
\label{Uk}
\begin{split}
U_k     &= \de{X^{k!}}^{\otimes k} \otimes Z \otimes \id^{\otimes n-k-2} \quad \text{for} \quad k<n-1, \\
U_{n-1} &= \de{X^{(n-1)!}}^{\otimes n-1},
\end{split}
\end{equation}
where we are using the convention that $A^{\otimes 0}=1$. For $n=4$, the set of unitaries defined by \eqref{Uk} reads
\begin{align} \nonumber
U_0 &= 	Z  \otimes \id\otimes\id , \\
U_1 &= 	X  \otimes Z  \otimes\id , \\\nonumber
U_2 &= 	X^2\otimes X^2\otimes Z  , \\\nonumber
U_3 &= 	X^6\otimes X^6\otimes X^6.
\end{align}
It is easy to check that the matrices generated according to these rules satisfy property \eqref{eq:pairwise}. A set of $n$ unitary matrices that satisfies property \eqref{eq:property_appendix} for an arbitrary $y$ can be found by repeating this construction with $\omega^y$ in place of $\omega$.

	Note that this construction is enough to show that there is an infinity of sets of unitaries satisfying property \eqref{eq:property_appendix}, since the choice of basis in the definition of the matrices $X$ and $Z$ in equation \eqref{eq:weylheisenberg} is arbitrary.

	This construction generates matrices with dimension $d=n!^{n-1}$, which therefore must act on $O(n^2\log n)$ qubits. One can also see that they can be implemented with a polynomial amount of elementary gates, since they are tensor products of a linear amount of $Z$ and $X^k$ matrices, that can themselves be implemented with a polynomial amount of elementary gates.
	
	One can also get constructions with the correct phases in lower dimension, which might be interesting for an experimental implementation. For example, for $n=3$, the following $6$-dimensional matrices also work:
	\begin{align} \nonumber
	U_0 &= Z, \\
	U_1 &= XZ, \\\nonumber
	U_2 &= X^2.
	\end{align}
	This construction saturates the lower bound $d\ge n!$ we presented in the main text. It is an interesting puzzle to find out whether there exists a construction with $d=n!$ for every $n$.

\section{Tolerating experimental error}\label{sec:tolerating}

	In an experimental implementation, no property $\textbf{P}_y$ can be satisfied exactly, so to run this algorithm one needs to be able to tolerate some deviations from it. To do this, notice that property $\textbf{P}_y$ is satisfied if and only if 
	\begin{equation}
	\label{b2}
	\frac{1}{n!^2d}\norm{\sum_{x=0}^{n!-1}\omega^{-xy}\Pi_x}^2_{HS} = 1,
	\end{equation}
	where $\norm{\cdot}_{HS}$ is the Hilbert-Schmidt norm. This is a simple consequence of the fact that this norm is defined through an inner product. We then reformulate the problem such that a set of unitaries satisfies the modified property $\textbf{P}_y'$ if 
	\begin{equation}
	\frac23 \le \frac{1}{n!^2d}\norm{\sum_{x=0}^{n!-1}\omega^{-xy}\Pi_x}^2_{HS} \le 1
	\end{equation}
	for a given $y \in \{0, \ldots, n!-1\}$. The problem is still to decide which property $\textbf{P}_y'$ the set of unitaries have, given the promise that they have one of them.

Note that, in this version of the problem, we cannot anymore use an arbitrary pure state $\ket{\psi}$ to perform the protocol, since it is not true anymore that the probabilities 
\[p_s = \frac{1}{n!^2}\norm{\sum_{x=0}^{n!-1}\omega^{-xs}\Pi_x\ket{\psi}}^2 \]
are independent of $\ket{\psi}$. In fact, it is possible to have a set of matrices such that the \textit{lhs} of Eq.~\eqref{b2} is arbitrarily close to $1$, while $p_y$ for some state $\ket{\psi}$ is equal to $0$. 

 Instead, we can use the maximally mixed state, since then the outcome probabilities $p_s$ become directly related to the Hilbert-Schmidt norm, as
	\begin{equation}
	 p_s = \frac{1}{n!^2d}\norm{\sum_{x=0}^{n!-1}\omega^{xs}\Pi_x}^2_{HS}
	\end{equation}
	and therefore the promise implies that $p_s \ge 2/3$ for $s=y$, and $p_s < 1/3$ for $s\neq y$. These two conditions are the ones necessary for a probabilistic algorithm to work; they imply that if we run the algorithm $k$ times and take as our guess for $y$ the most common answer, the probability of making a mistake goes down exponentially with $k$.

\section{\texorpdfstring{Implementing the $n$-\textsc{switch} gate in the quantum circuit model}{Implementing the n-switch gate in the quantum circuit model}}\label{sec:switchbqp}

Here we describe how to simulate the $n$-\textsc{switch} gate in the quantum circuit model. The simulation is based on the circuit presented on Ref.~\cite{colnaghi11}, with the difference that our scheme can be used for quantum (and not only classical) control of the order.

  In the main text we described the control system $\ket{C}$ as encoding the permutation to be applied simply as a state running from $\ket{0}$ to $\ket{n!-1}$. Here we shall use a more convenient representation, expressing $\ket{C} = \ket{C_1}\ldots\ket{C_n}$, where each $\ket{C_k}$ runs from $\ket{0}$ to $\ket{n-1}$, and indicates the unitary to be applied in the position $k$. For example, to encode the permutation that first applies $U_1$, then $U_0$, and then $U_2$, we shall write $\ket{C} = \ket{102}$. This representation requires $n \lceil \log_2 n \rceil$ qubits, a small overhead over the $\lceil \log_2 n! \rceil$ qubits that are necessary to encode a permutation of $n$ elements. We also remark that the conversion between these two representations can be done on a classical computer in polynomial time, and therefore we shall note it no further.

The circuit consists of a composition of $n$ instances of the following element, where $k$ goes from $1$ to $n$:
\[
\Qcircuit @C=0.5em @R=0.9em {
\lstick{\ket{C_k}}     & \ctrl{1}           & \qw 	& \ctrl{1}          & \qw  \\
\lstick{\ket{\psi}}  & \multigate{3}{S}   & \qw 	& \multigate{3}{S}  & \qw  \\
\lstick{\ket{a_0}}   & \ghost{S}	  & \gate{U_0} 	&  \ghost{S}	    & \qw  \\
\lstick{\vdots\phantom{m}}   & \pureghost{S}	  & \vdots 	&  \pureghost{S}        & 	  \\
\lstick{\ket{a_{n-1}}}   & \ghost{S}       	  & \gate{U_{n-1}}	&  \ghost{S}        & \qw	       
}\]
The $\ket{a_i}$ are ancillæ, and $S$ is a gate that swaps $\ket{\psi}$ with the ancilla $\ket{a_i}$ controlled on $\ket{C_k}=\ket{i}$, leaving the other ancillæ invariant. This gate clearly can be implemented with a linear amount of elementary gates, so we shall not discuss its implementation. Note that each element contains $n$ unitaries; therefore, since we need $n$ of these elements to implement the $n$-\textsc{switch}, the total number of queries in this implementation is $n^2$.

For example, using this construction for $n=3$ gives us the circuit
\[
\Qcircuit @C=0.5em @R=0.9em {
\lstick{\ket{C_3}}     & \qw		    & \qw		& \qw			 & \qw
		       & \qw		    & \qw		& \qw			 & \qw
		       & \ctrl{3}           & \qw 	        & \ctrl{3}               & \qw  \\
\lstick{\ket{C_2}}     & \qw		    & \qw		& \qw			 & \qw
		       & \ctrl{2}           & \qw 	        & \ctrl{2}               & \qw
		       & \qw		    & \qw		& \qw			 & \qw  \\
\lstick{\ket{C_1}}     & \ctrl{1}           & \qw 	        & \ctrl{1}               & \qw  
		       & \qw		    & \qw		& \qw			 & \qw
		       & \qw		    & \qw		& \qw			 & \qw  \\
\lstick{\ket{\psi}}    & \multigate{3}{S}   & \qw 	        & \multigate{3}{S}  & \qw  
		       & \multigate{3}{S}   & \qw 	        & \multigate{3}{S}  & \qw
		       & \multigate{3}{S}   & \qw 	        & \multigate{3}{S}  & \qw  \\
\lstick{\ket{a_0}}     & \ghost{S}	    & \gate{U_0} 	&  \ghost{S}	 & \qw  
		       & \ghost{S}	    & \gate{U_0} 	&  \ghost{S}	 & \qw  
		       & \ghost{S}	    & \gate{U_0} 	&  \ghost{S}	 & \qw  \\
\lstick{\ket{a_1}}     & \ghost{S}	    & \gate{U_1} 	&  \ghost{S}        & \qw  
		       & \ghost{S}	    & \gate{U_1} 	&  \ghost{S}        & \qw  
		       & \ghost{S}	    & \gate{U_1} 	&  \ghost{S}        & \qw  \\
\lstick{\ket{a_2}}     & \ghost{S}          & \gate{U_2}	&  \ghost{S}        & \qw
		       & \ghost{S}          & \gate{U_2}	&  \ghost{S}        & \qw	       
		       & \ghost{S}          & \gate{U_2}	&  \ghost{S}        & \qw	       
}\]

Note that when each unitary is applied only once to the target system, \ie when $\ket{C}$ encodes a permutation or a superposition of permutations, then the ancillæ disentangle from the control system at the end of the circuit (since $U_i$ is applied $n-1$ times on $\ket{a_i}$, independently of $\ket{C}$). This is not the case when $\ket{C}$ encodes a more general sequence of unitaries, thus this circuit cannot be used to implement quantum control of arbitrary sequences of $n$ unitaries\footnote{The ancillæ also disentangle for sequences in which each unitary is applied to the target a fixed number of times. For example, when $\ket{C}$ is in a superposition of $\ket{002}$, $\ket{020}$, and $\ket{200}$ the ancillæ still disentangle (in each case $U_0$ is applied twice and $U_2$ once), but not when it is a superposition of $\ket{002}$ and $\ket{021}$.}.

If the time used by the swap gates is neglected, this circuit has running time $n$. It is easy to see that a circuit with this running time cannot use less than $n^2$ queries: in order to be possible to implement all permutations, each of the $n$ unitaries must be available in each of the $n$ time steps. It is however possible to reduce the number of queries at the expense of the running time (for the 2-\textsc{switch}, an implementation with one query to $U_0$ and two queries to $U_1$ is possible, see Ref.~\cite{chiribella12}.) As shown in the main text, no implementation with less than $\Omega(n^2)$ queries is possible. Note that this coincides with the definition of ``gate uses'' introduced in the main text.

\section{Relationship with other quantum algorithms}\label{sec:other_algorithms}

  We would like to understand the relationship of our algorithm with other known classes of quantum algorithms; in particular, there are some similarities between our algorithm and those that solve the (abelian) hidden subgroup problem, and we wanted to explore how deep they are.
  
  In the hidden subgroup problem one is given a group $G$ and needs to find (a set of generators for) a hidden subgroup $H$, by using a function $f(g)$ that is constant in each coset of $H$, and different on different cosets. One is given access to $f$ via a black-box unitary that does the mapping
\[ U\ket{x}\ket{y} = \ket{x}\ket{y + f(x)}.\]
In our problem, one is given as input a set of unitaries $\{U_j\}_{j=0}^{n-1}$, with the promise that its permutations $\Pi_x$ act as
\[ S_n\ket{x}\ket{\psi} = \ket{x}\Pi_x\ket{\psi} = \omega^{xy}\ket{x}\Pi_0\ket{\psi}. \]
A first difference is that our unitaries act by applying a phase, instead of shifting the state of the ancilla. But if one nevertheless considers the function $f(x) = \omega^{xy}$ one can see that it obeys the promise of the hidden subgroup problem: it is a periodic function with period $r= n!/\gcd(y_0,n!)$, and if one takes the group as $G = Z_{n!}$ with addition modulo $n!$, and $H = \{0, r, 2r, \ldots \}$, then $f(x)$ is constant in each coset of $H$ and distinct for distinct cosets. Moreover, our algorithm finds $y$ and therefore $r$, solving the hidden subgroup problem for this function.

Can we push this analogy further? More specifically, can we use our algorithm to find the period of any function $f'(x)$ that is constant on each coset of the hidden subgroup and different for different cosets? We are going to show that this is not the case. Let then $f'(x) = \omega^{g(x)}$ be such a function. We start in the state
		\[ \ket{C}\ket{\psi} = \frac{1}{\sqrt{n!}}\sum_{x=0}^{n!-1}\ket{x}\ket{\psi}, \]
		and apply the switch:
		\[ S_n\ket{C}\ket{\psi} = \frac{1}{\sqrt{n!}}\sum_{x=0}^{n!-1}\ket{x}\Pi_x\ket{\psi}  = \frac{1}{\sqrt{n!}}\sum_{x=0}^{n!-1}\omega^{g(x)}\ket{x}\Pi_0\ket{\psi}. \]
		Now we split the sum into the subgroup $H$ and its cosets:
		\begin{align*}
		S_n\ket{C}\ket{\psi} &=\frac{1}{\sqrt{n!}}\sum_{x=0}^{n!-1}\omega^{g(x)}\ket{x}\Pi_0\ket{\psi} \\
				      &= \frac{1}{\sqrt{n!}}\sum_{l=0}^{r-1}\sum_{a=0}^{\frac Nr-1}\omega^{g(l+ar)}\ket{l+ar}\Pi_0\ket{\psi}.
		\end{align*}
		Now we use the fact that $g(x+r) = g(x)$,
		\[ S_n\ket{C}\ket{\psi} = \frac{1}{\sqrt{n!}}\sum_{l=0}^{r-1}\sum_{a=0}^{\frac Nr-1}\omega^{g(l)}\ket{l+ar}\Pi_0\ket{\psi}, \] 
		and apply the inverse Fourier transform on the first register:
		\begin{align*}
		\mathcal{F}_{n!}S_n\ket{C}\ket{\psi} &= \frac{1}{n!}\sum_{s=0}^{n!-1}\sum_{l=0}^{r-1}\sum_{a=0}^{\frac {n!}r-1}\omega^{g(l)-y(l+ar)}\ket{s}\Pi_0\ket{\psi} \\
						      &= \frac1r\sum_{k=0}^{r-1} \de{\sum_{l=0}^{r-1}\omega^{g(l)-k\frac {n!}rl}} \ket{k{\textstyle\frac {n!}r}}\Pi_0\ket{\psi}
		\end{align*}
		The algorithm ends by measuring the first register in the computational basis. At this point, it is useful to compare this state to the one you would get if you were applying the period-finding algorithm. With it, the state of the first register would be \[\frac1{\sqrt r}\sum_{k=0}^{r-1} \ket{k{\textstyle\frac {n!}r}}, \]
		a \textit{uniform} superposition over the multiples of $\frac {n!}r$, and a constant amount of repetitions of the algorithm would be enough to determine $r$ with high probability.
		
		Instead, because in our case the superposition is not uniform, one needs an exponential amount of repetitions in the worst case. To see that, consider the case when $g(l) = l \mod n!$. Then the probability of obtaining $\ket{0}$ when measuring the first register is
		\[
		\frac{\sin^2\de{\frac{\pi r} {n!}}}{r^2\sin^2\de{\frac{\pi} {n!}}},
		\] 
		and this means that, for constant $r$, one would need to make $O(n!^2)$ measurements to get an outcome different than $\ket{0}$ and any information at all about the period $r$.
				
		For this reason, it seems that the analogy between our algorithm and the period-finding algorithm (and, therefore, those solving the hidden subgroup problem) cannot be pushed far. We think the true connection is only at the more basic level that they are both applications of the quantum Fourier transform.

\section{\texorpdfstring{Decomposition of the $n$-\textsc{router}}{Decomposition of the n-router}}\label{sec:router}
We show how an $n$-\textsc{router} can be constructed using a polynomial number of elementary resources, each equivalent to a polarizing beam splitter (PBS). Let $\kett{x}$ be a basis element of the control system, and $\kett{j}_{\hbox{in}}$ denote the $j$-th input mode to the router, with $j=0,\dots, n-1$ (for simplicity, we don't write explicitly the target system). Then the $n$-\textsc{router} performs the transformation $\kett{x}\kett{j}_{\hbox{in}}\mapsto \kett{x}\kett{\sigma_x(j)}_{\hbox{out}}$, where $\kett{\sigma_x(j)}_{\hbox{out}}$ is the $\sigma_x(j)$-th output mode. In order to decompose this gate in elementary ones, we first express the control variable in terms of its factoradic representation, introduced in Section \ref{sec:fabio}: $x\leftrightarrow \left(a_{n-1},\dots,a_1\right)$, with $0\leq a_k \leq k$. Then, we represent each coefficient $a_k$ using $k$ bits\footnote{This encoding is clearly not optimal, since it requires $O(n^2)$ bits to encode the $n!$ permutations, instead of the minimal $O(n\log n)$. We will however not invest more time to optimize this aspect of the problem.}
$b_{k\,1},\dots,b_{k\,k}$, with $b_{k\,j}=1$ for $j \le a_k$ and $b_{k\,j}=0$ for $j > a_k$. For example, the values of $a_3$ are represented as
\begin{center}
	\begin{tabular}{r l l l l l }
	  	    $a_3 =$ &  & 0 & 1 & 2 & 3 \\ \hline
      $b_{3\,1}=$ &  & 0 & 1 & 1 & 1 \\ 
			$b_{3\,2}=$ &  & 0 & 0 & 1 & 1 \\ 
			$b_{3\,3}=$ &  & 0 & 0 & 0 & 1 \\ 
	\end{tabular}
	\label{tab:}
\end{center}
In this way, we can encode the control quantum system in $\frac{n(n-1)}{2}$ qubits, $\kett{x}\rightarrow\bigotimes_{k=1}^{n-1}\bigotimes_{j=1}^k\kett{b_{k\,j}}$. Now we can construct the $n$-\textsc{router} using a controlled binary swap of modes for each control qubit $\kett{b_{k\,j}}$. Recall that an arbitrary permutation $\kett{j}_{\hbox{in}}\mapsto \kett{\sigma_x(j)}_{\hbox{out}}$ is obtained shifting each mode $k$ ``to the right'' by a number $a_k$ of positions, \ie applying the unitary $\kett{k}_{\hbox{in}}\mapsto \kett{k-a_k}_{\hbox{out}}$, $\kett{j}_{\hbox{in}}\mapsto \kett{j+1}_{\hbox{out}}$ for $k-a_k\leq j<k$. The idea is to decompose this shift in swaps between neighboring modes, with each swap controlled by one control qubit.
Explicitly, the controlled mode-swaps are defined as 
\begin{align}
\kett{b_{k\,j}}\kett{k-j}_{\hbox{in}}   \mapsto & \kett{b_{k\,j}}\kett{k-j+b_{k\,j}}_{\hbox{out}} \\\nonumber
\kett{b_{k\,j}}\kett{k-j+1}_{\hbox{in}} \mapsto & \kett{b_{k\,j}}\kett{k-j+1-b_{k\,j}}_{\hbox{out}},
\end{align}
and identity for the other modes. The $n$-\textsc{router} is then obtained by first applying the mode-swap controlled by $\kett{b_{1\,1}}$, then the one controlled by $\kett{b_{2\,1}}$ followed by the one controlled by $\kett{b_{2\,2}}$ and so on, applying successively each mode-swap controlled by $\kett{b_{k\,j}}$ increasing $k$ and, for each $k$, increasing $j$.

As an example, consider a permutation identified by the single non-vanishing factoradic coefficient $a_3=2$. The corresponding control qubits are then in the states $\kett{b_{3\,1}=1}\kett{b_{3\,2}=1}\kett{b_{3\,3}=0}$. First we apply the swap between modes $3$ and $2$ controlled on  $\kett{b_{3\,1}}$. Since $b_{3\,1}=1$ this results in the mode-swap $\ket{3}\leftrightarrow\ket{2}$.  Then, since $b_{3\,2}=1$, the modes $2$ and $1$ are swapped. Composing the two swaps, we obtain the transformation $\ket{3}\rightarrow\ket{1}$, $\ket{1}\rightarrow\ket{2}$, $\ket{2}\rightarrow\ket{3}$. Since $b_{3\,3}=0$, the last mode-swap is not applied, the result of the two mode-swaps is therefore the shift of mode $3$ by $a_3=2$ positions to the right. By composing this procedure for an arbitrary set of coefficients $\left(a_{n-1},\dots,a_1\right)$, the corresponding permutation of modes is realized.

The building block of this construction is the controlled swap of two modes, which is essentially equivalent to a PBS (for a PBS, the control qubit is the photon polarization). We stress that this element does not correspond to any traditional elementary gate of a quantum circuit, and it should thus be considered as a new elementary resource.  

In Ref.~\cite{colnaghi11}, a different approach was proposed for the scalable realization of the $n$-\textsc{switch}: a construction was proposed that realizes the $n$-\textsc{switch} using $O(n^2)$ 2-\textsc{switch}es. This construction, however, is not applicable to our case, since our definitions differ slightly\footnote{According to the definition from~\cite{colnaghi11}, the $n$-\textsc{switch} orders the $n$ unitaries according to the prescribed permutation, but it does not directly compose them to each other. It is thus possible to plug additional elements between any pairs of unitaries, making the construction of the $n$-\textsc{switch} from 2-\textsc{switch}es possible. The 2-\textsc{switch} of~\cite{colnaghi11} can be reproduced by a 3-\textsc{switch} as defined here, with the prescription that only the first and last position are swapped, while the position in the middle is fixed and can be used either as an identity or to plug input and output of another switch.}. Furthermore, the construction based on the $n$-\textsc{router} element is more directly related to the interferometric implementation proposed in the main text.

\end{document}